\documentclass[preprint]{raa}            
\usepackage{ifpdf}
\usepackage{graphicx,times}             
\usepackage{natbib}
\usepackage{amssymb,amsmath}
\bibpunct{(}{)}{;}{a}{}{,}

\usepackage[pagebackref=true]{hyperref}

\begin{document}

  \title{Constraining the Neutral Hydrogen Fraction from SKA Simulated Observation using a Double-Gaussian Decomposition Technique}

   \volnopage{Vol.0 (20xx) No.0, 000--000}      
   \setcounter{page}{1}          

   \author{Jiajun Zhang
      \inst{1,2}, Huanyuan Shan \inst{1,2}
   }
   

   \institute{1 Shanghai Astronomical Observatory, Chinese Academy of Sciences, China;\\
   2 Key Laboratory of Radio Astronomy and Technology, Chinese Academy of Sciences, 20A Datun Road, Chaoyang District, Beijing 100101, China\\
   {\it jjzhang@shao.ac.cn}\\}

\vs\no
   {\small Received 20xx month day; accepted 20xx month day}

\abstract{The Epoch of Reionization (EoR) is a unique phase in cosmic history, marked by the ionization of neutral hydrogen by the first luminous sources. The global neutral hydrogen fraction ($x_{HI}$) is a key observable for probing this era. This paper presents a novel, statistically robust method to extract the evolution of $x_{HI}$ from the challenging noise-dominated data from the Square Kilometre Array (SKA) Data Challenge 3b. Our approach is based on a key physical insight: the pixel value distribution in SKA intensity maps is a mixture of signals from ionized and neutral regions. We model this distribution as a superposition of two Gaussian components—one fixed at zero representing noise and ionized bubbles, and a second, offset Gaussian tracing the neutral hydrogen signal. We perform this decomposition on data grouped into three redshift bins. The double-Gaussian model provides an excellent fit to the pixel histogram data. The derived $x_{HI}$ values show a clear decreasing trend across the three redshift bins, consistent with a progressing reionization process. And the results are consistent with the provided simulation data. This method offers a powerful, model-independent, and fully interpretable way for measuring $x_{HI}$ from 21 cm data, demonstrating significant potential for application to future SKA observations.}
\keywords{cosmology: observations --- dark ages, reionization, first stars --- techniques: image processing --- methods: data analysis --- methods: statistical}

   \authorrunning{Jiajun Zhang}            
   \titlerunning{HI fraction inference}  

   \maketitle

%
%

\section{Introduction}
Within several million years following the Big Bang, the complex architecture of the cosmic web started to assemble as a consequence of dark matter collapsing under its own gravity. During this era, the initial stellar and galactic systems emerged, whose radiative output progressively ionized the neutral baryonic gas. This phase, termed cosmic reionization, constituted a fundamental transition in the Universe's history, paving the way for the rich variety of structures and astrophysical objects seen today (see, e.g., \citep{Morales10} for a detailed overview).

To this day, the cosmic Dark Ages and the subsequent reionization era represent compelling and still largely uncharted chapters in our knowledge of post-Cosmic Microwave Background (CMB) evolution. Deeper exploration and progress in observational methods—such as the deployment of advanced telescope arrays and high-precision measurement capabilities—are crucial for understanding the young universe.

The 21 cm spectral line, originating from the hyperfine splitting of neutral hydrogen's ground state, provides a vital observational window into the Universe throughout the reionization epoch \citep{Pritchard12}. This signal, often called the EoR signal, has been redshifted into the low-frequency radio domain, typically observed between approximately 50 MHz and 200 MHz. To access this signal, several radio interferometric facilities have been developed or proposed, such as the Square Kilometre Array (SKA, \citep{SKA09}), the Hydrogen Epoch of Reionization Array (HERA, \citep{HERA22}), the Low-Frequency Array (LOFAR, \citep{LOFAR13}), the Murchison Widefield Array (MWA, \citep{MWAII18}), the 21 Centimeter Array (21CMA, \citep{21CMA09}), and the Precision Array for Probing the Epoch of Reionization (PAPER, \citep{PAPER07}).

Of these instruments, the SKA-Low array represents one of the most ambitious initiatives, with physical construction now underway. It is expected to achieve exceptional sensitivity and angular resolution, positioning it as a leading facility for the probable detection of the EoR signal. SKA Observatory launched a series of data challenges to attract scientists worldwide to participate in detecting the EoR signal. The data challenge that calls for scientists to propose methods to remove foreground has concluded \citep{sdc3a}\footnote{https://sdc3.skao.int/challenges/foregrounds}. Another challenge is to estimate the neutral hydrogen fraction $x_{HI}$ from simulations. Several methods were proposed to estimate $x_{HI}$ from EoR 21cm power spectrum\citep{Liu2025GPR, Gonzalez2025BNRE}. In this paper, we would like to propose a fast method to calculate the $x_{HI}$ from imaging data with good interpretability.

We present the data and methodology in Sec.~\ref{sec:data}. The Results are presented in Sec.~\ref{sec:result}. Finally, we provide the conclusion and discussion in Sec.~\ref{sec:conclusion}.

\section{Data and Methodology}
\label{sec:data}
\subsection{SKA Data Challenge 3b Data}
SKA Observatory launched the second phase of Data Challenge 3 (SDC3b) in late 2024 and concluded the results collection at the end of May 2025. This data challenge provided both brightness temperature power spectrum data and imaging data generated from simulations that replicate real observations with SKA-Low. The imaging data include the 21 cm signal from the Epoch of Reionization, residual foreground contamination, and noise corresponding to 1000 hours of observation. The data cover the full frequency range from 151 MHz to 196 MHz, divided into 450 frequency bands. Each band contains an image with $2048 \times 2048$ pixels. The data are provided in FITS format and are available on the SDC3b website\footnote{https://sdc3.skao.int/challenges/inference}. The goal of the inference task is to estimate the neutral hydrogen fraction ($x_{\text{HI}}$) at three frequency bands and to provide the associated uncertainty. The cosmological and astrophysical model parameters are model-dependent. Estimating the evolution of the neutral fraction serves as a fundamental basis for model-independent analysis of the Epoch of Reionization.
\subsection{Data Pre-processing}
\begin{figure}
    \centering
    \includegraphics[width=0.95\linewidth]{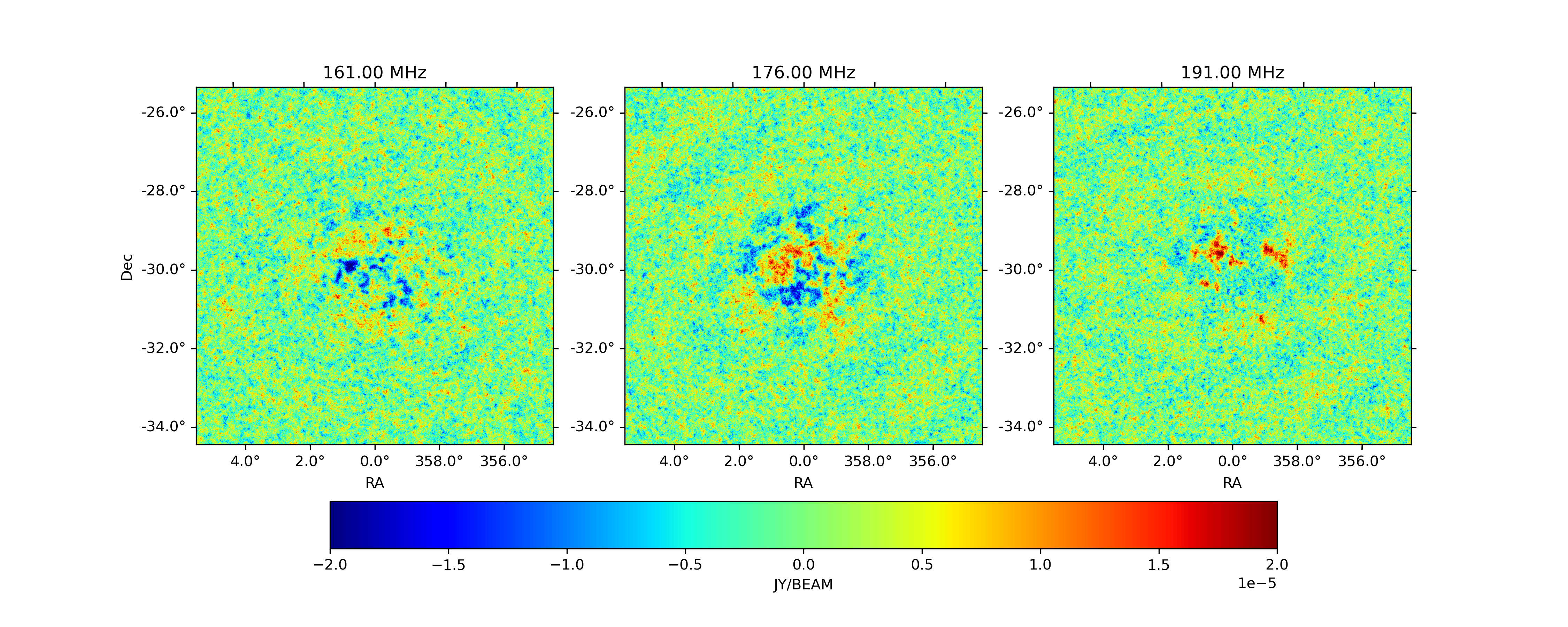}
    \caption{Three example images from SDC3b, showing the luminosity distribution in the unit of Jy/beam, at 161MHz, 176MHz, and 191MHz. Structured patterns can be seen in the central region.}
    \label{fig:image}
\end{figure}

Figure~\ref{fig:image} illustrates the images at 161 MHz, 176 MHz, and 191 MHz, with the color scale representing luminosity. Two key features are evident: first, the highest values are concentrated in the central region rather than the outskirts; second, while the pixel values in the outskirts appear randomly distributed, the central region exhibits coherent structures. As these images are generated from simulated interferometric data lacking a monopole component, the mean pixel value is approximately zero. The primary synthesized beam, formed by the SKA-Low antennas prior to interferometry, is represented by the central region of the images. It is therefore advisable to extract this central region prior to analysis. We extract a circular region with a radius of 400 pixels, applying the same area across all frequency channels.

\subsection{The Physical Basis for the Double-Gaussian Model}
Under ideal conditions, free from foreground contamination and thermal noise, the 21cm brightness temperature map would directly trace the neutral hydrogen distribution. In this scenario, ionized regions would exhibit no signal beyond the Cosmic Microwave Background (CMB). A reasonable assumption is that the underlying 21cm signal follows a Gaussian distribution with a non-zero mean, reflecting the large-scale structure. In contrast, the combined effect of foreground removal residuals and thermal noise is also commonly modeled as a Gaussian distribution, but one centered at zero, as these components do not correlate with the cosmological signal.

Consider two independent groups of pixels:
\begin{itemize}
    \item All zeros, represent the reionized pixels.
    \item Values drawn from a Gaussian distribution $Y \sim \mathcal{N}(\mu_y, \sigma_y^2)$, represent the EoR signal pixels.
\end{itemize}
Both groups are added to a Gaussian random variable $X \sim \mathcal{N}(0, \sigma_x^2)$, which is independent of $Y$, represents the noise component (foreground removal residual and thermal noise). After adding $X$:
\begin{itemize}
    \item For reionized pixels, the values become $X + 0 = X$, so the distribution is $N_1 \sim \mathcal{N}(0, \sigma_x^2)$.
    \item For EoR signal pixels, the values become $X + Y$, and since $X$ and $Y$ are independent, the distribution is $N_2 \sim \mathcal{N}(\mu_y, \sigma_x^2 + \sigma_y^2)$.
\end{itemize}

$N_1$ is the number of reionized pixels and $N_2$ is the number of EoR signal pixels. The total number of pixels is $N = N_1 + N_2$. The overall data set is a mixture of two Gaussian distributions:
\begin{align}
    p(z) = \pi_1 \cdot \mathcal{N}(z \mid 0, \sigma_x^2) + \pi_2 \cdot \mathcal{N}(z \mid \mu_y, \sigma_x^2 + \sigma_y^2)
\end{align}
where $\pi_1 = \frac{N_1}{N}$ and $\pi_2 = \frac{N_2}{N}$ are the mixing weights, with $\pi_1 + \pi_2 = 1$.

Therefore, we can try to fit the pixel values in the image by a double-Gaussian model as follows:
\begin{equation}
N(x) = A_\mathrm{n} \cdot \exp \left[ -\frac{1}{2} \left( \frac{x}{\sigma_\mathrm{n}} \right)^2 \right] + A_\mathrm{s} \cdot \exp \left[ -\frac{1}{2} \left( \frac{x - \mu_\mathrm{s}}{\sigma_\mathrm{s}} \right)^2 \right],
\end{equation}
where $N(x)$ represents the pixel count distribution for different pixel value $x$, $A_\mathrm{n}$ and $\sigma_\mathrm{n}$ represent the weight and standard deviation for the noise part (foreground removal residual and thermal noise), $A_\mathrm{s}$, $\mu_\mathrm{s}$ and $\sigma_\mathrm{s}$ represent the weight, mean value and standard deviation for the 21cm signal part.

\subsection{Analysis Pipeline}
After fitting, the area under the nonzero-centered Gaussian part represents the number of EoR signal pixels. Divided by the total area under that double-Gaussian,  we get the very first estimate of neutral hydrogen fraction $x_{HI}$. Finally, we should consider that the evolution of $x_{HI}$ being smooth, the bins are chosen not exactly at the frequency required by SDC3b, and the groups are correlated as well. The joint probability distribution is derived through Gaussian Process regression on the binned $x_{HI}$ estimates, modeling the covariance structure across redshift bins to capture the smooth evolution of reionization while accounting for measurement errors.

Given the above consideration, the pipeline can be constructed in the following steps:
\begin{itemize}
    \item Grouping the data into three redshift bins
    \item Pick out the central spherical area with a radius of 400 pixels
    \item Construct the pixel value histograms for each redshift bin
    \item Performing the non-linear least-squares fit with the double Gaussian model
    \item Calculating $x_{HI}$ from the ratio of the integrated areas of the Gaussian components
    \item Propagating fitting errors to estimate uncertainties on $x_{HI}$
    \item Calculate the final probability distribution using Gaussian Process regression
\end{itemize}

\section{Results}
\label{sec:result}
\begin{figure}
    \centering
    \includegraphics[width=0.95\linewidth]{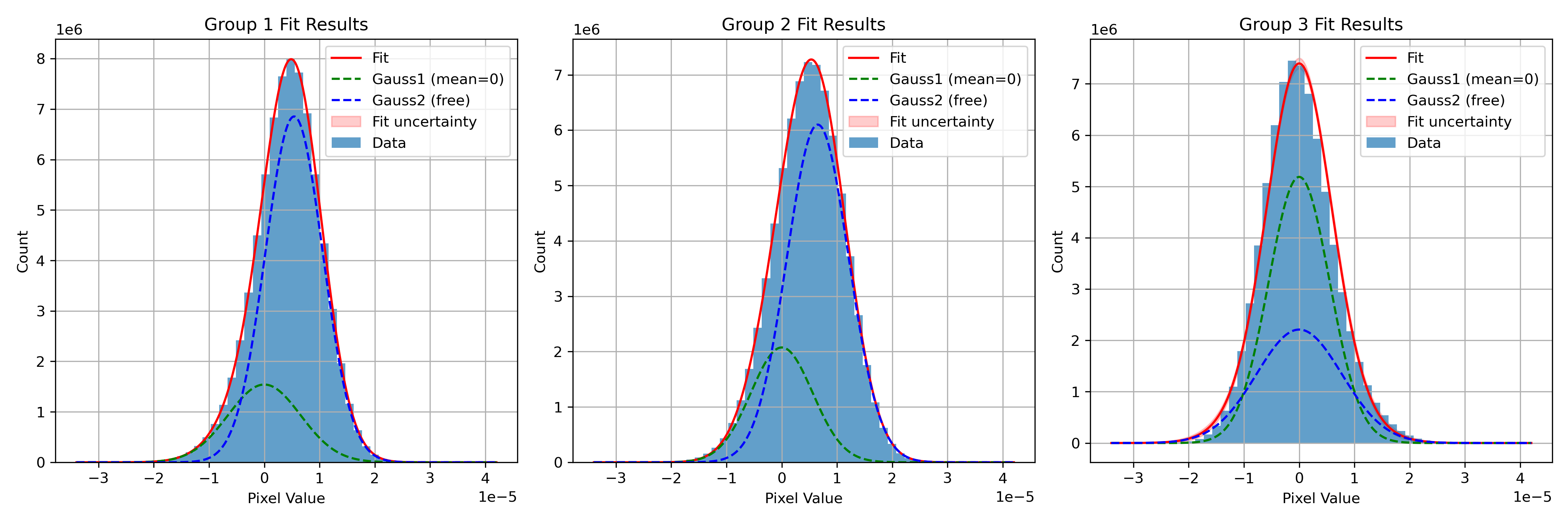}
    \caption{The histogram and double-Gaussian model fitting results for the pixel values in the images. Group 1 contains 151MHz-166MHz, Group 2 contains 166MHz-181MHz, Group 3 contains 181MHz-196MHz. The green dashed line represents the contribution from foreground removal residual and thermal noise. The blue dashed line represents the contribution from 21cm signal.}
    \label{fig:hist}
\end{figure}

\begin{figure}
    \centering
    \includegraphics[width=0.95\linewidth]{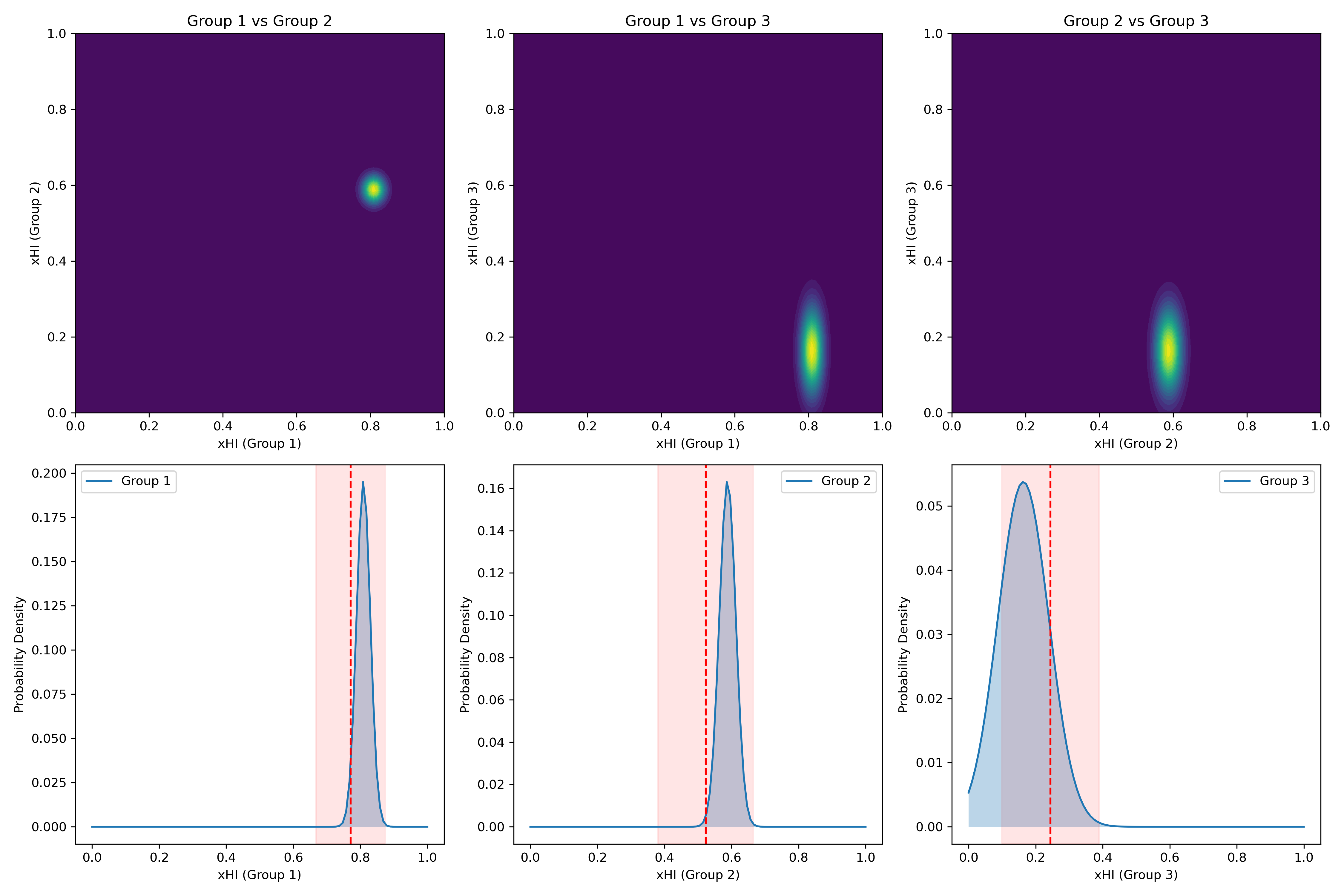}
    \caption{The top panels show the two-dimensional joint probability for 3 groups of frequency bins. The bottom panels show the one-dimensional probability inference results using our method. The red dashed line is the central value answer, and the shaded red region is the $1\sigma$ range, obtained from the SDC3b website.}
    \label{fig:inference}
\end{figure}

We have grouped the data of images into three frequency bins, 151MHz-166MHz are grouped as Group 1, 166MHz-181MHz are grouped as Group 2, and 181MHz-196MHz are grouped as Group 3. The histogram of these three groups and the double-Gaussian model fitting are shown in Fig.~\ref{fig:hist}. Note that the green dashed line represents the noise part, which contributes more and more with higher and higher frequency. This is expected because it reflects the larger and larger reionization regions. The fitting results are shown in Tab.~\ref{tab:gaussian_fits}. And the EoR signal pixel area ratio for Group 1 is $0.7789 \pm 0.0059$, for Group 2 is $0.7466 \pm 0.0216$, and for Group 3 is $0.3741 \pm 0.0335$.
\begin{table}[htbp]
\centering
\caption{Fitting results for three groups of data using the double-Gaussian model. $A_n$ and $\sigma_n$ represent the amplitude and standard deviation of the Gaussian with fixed mean=0 (reionized pixels), while $A_s$, $\mu_s$, and $\sigma_s$ represent the amplitude, mean, and standard deviation of the free Gaussian (EoR signal pixels).}
\label{tab:gaussian_fits}
\begin{tabular}{|l|c|c|c|c|c|}
\hline
Group &  $A_n$ & $\sigma_n$ & $A_s$ & $\mu_s$ & $\sigma_s$ \\
Group 1 &  $1.541\times10^{6}$ & $6.560\times10^{-6}$ & $6.854\times10^{6}$ & $5.386\times10^{-6}$ & $5.194\times10^{-6}$ \\
Group 2 &  $2.074\times10^{6}$ & $5.579\times10^{-6}$ & $6.100\times10^{6}$ & $6.521\times10^{-6}$ & $5.590\times10^{-6}$ \\
Group 3 &  $5.186\times10^{6}$ & $5.519\times10^{-6}$ & $2.211\times10^{6}$ & $0.000$ & $7.736\times10^{-6}$ \\
\hline
\end{tabular}
\end{table}
Following the pipeline described before, we provide the inference results for these three frequency bins. After Gaussian Process regression, the estimated value of $x_{HI}$ for Group 1 is $0.8092\pm0.021$, for Group 2 is $0.5883\pm0.025$, and for Group 3 is $0.1633\pm0.076$.  We illustrate the two-dimensional joint probability results and the one-dimensional probability results in Fig.~\ref{fig:inference}. The one-dimensional probability answer provided by SKA SDC3b is shown in the shaded red region, the red dashed line shows the central value, and the shaded area shows the $1\sigma$ region. The inference probability results using our methods are shown in the shaded blue region. It is clear that our method can provide a fast and accurate inference for the $x_{HI}$ value, and the process is highly interpretable. Our method is the YEYE team method submitted to SDC3b, with a different choice of binning scheme, resulting in a more precise probability distribution.

\section{Conclusion and Discussion}
\label{sec:conclusion}
In this paper, we have successfully developed a method to estimate $x_{HI}$ from SKA-like data using a double-Gaussian decomposition, revealing a plausible reionization history. Our method is fast, accurate, and easy to understand. We start from the guessing process when looking at the data images. Turns out that the initial guess and the implementation of the double-Gaussian decomposition method are successful.  

We acknowledged that the recent development in machine learning methods application in the field of EoR detection is successful. However, the poor interpretability of the machine learning methods prevents us from understanding the physics behind. Our method looks simple, but useful, and originated from human intuition. It is interesting to understand how human intuition works, transfer it into logical steps and apply it to our applications.

\section{Acknowledgments}
This work is supported by the Ministry of Science and Technology of China (grant No. 2020SKA0110100) and the National Key R\&D Program of China No. 2022YFF0504300. This work makes use of data from the SKA Data Challenge 3b. This work is supported by the Specialized Research Fund for State Key Laboratory of Radio Astronomy and Technology.

\bibliographystyle{raa}
\bibliography{yeye}

\end{document}